# Quasi-Ballistic Thermal Transport Across MoS$_2$ Thin Films

*Aditya Sood*[1,2,†,‡,*], *Feng Xiong*[3,‡], *Shunda Chen*[4], *Ramez Cheaito*[2], *Feifei Lian*[1,¶], *Mehdi Asheghi*[2], *Yi Cui*[5,6], *Davide Donadio*[4,7], *Kenneth E. Goodson*[2,5], *Eric Pop*[1,5,8,*]

[1]Department of Electrical Engineering, Stanford University, Stanford, CA 94305, USA. [2]Department of Mechanical Engineering, Stanford University, Stanford, CA 94305, USA. [3]Department of Electrical and Computer Engineering, University of Pittsburgh, Pittsburgh, PA 15261, USA. [4]Department of Chemistry, University of California, Davis, CA 95616, USA. [5]Department of Materials Science and Engineering, Stanford University, Stanford, CA 94305, USA. [6]Stanford Institute for Materials and Energy Sciences, SLAC National Accelerator Laboratory, Menlo Park, CA 94025, USA. [7]Ikerbasque, Basque Foundation for Science, E-48011 Bilbao, Spain. [8]Precourt Institute for Energy, Stanford University, Stanford, CA 94305, USA.

ABSTRACT: Layered two-dimensional (2D) materials have highly anisotropic thermal properties between the in-plane and cross-plane directions. Conventionally, it is thought that cross-plane thermal conductivities ($\kappa_z$) are low, and therefore *c*-axis phonon mean free paths (MFPs) are small. Here, we measure $\kappa_z$ across MoS$_2$ films of varying thickness (20 to 240 nm) and uncover evidence of very long *c*-axis phonon MFPs at room temperature in these layered semiconductors. Experimental data obtained using time-domain thermoreflectance (TDTR) are in good agreement with first-principles density functional theory (DFT). These calculations suggest that ~50% of the heat is carried by phonons with MFP >200 nm, exceeding kinetic theory estimates by nearly two orders of magnitude. Because of quasi-ballistic effects, the $\kappa_z$ of nanometer-thin films of MoS$_2$ scales with their thickness and the volumetric thermal resistance asymptotes to a non-zero value, ~10 m$^2$KGW$^{-1}$. This contributes as much as 30% to the *total* thermal resistance of a 20 nm thick film, the rest being limited by thermal interface resistance with the SiO$_2$ substrate and top-side aluminum transducer. These findings are essential for understanding heat flow across nanometer-thin films of MoS$_2$ for optoelectronic and thermoelectric applications.

**Introduction**

Two-dimensional (2D) van der Waals (vdW) layered solids have highly unusual thermal transport properties due to their unique crystal structure. While atoms within a layer are bonded covalently, adjacent layers are coupled via weak vdW interactions. This leads to a strong anisotropy in thermal conductivity, with the in-plane (along the layers) conductivity $\kappa_r$ being significantly higher than the cross-plane (across the layers, or along the c-axis) conductivity $\kappa_z$. For example, in bulk graphite, h-BN, and MoS$_2$, anisotropy ratios ($\kappa_r/\kappa_z$) as high as ~300, 200, and 50, respectively, have been reported at room temperature.[1–3] Owing to their high $\kappa_r$, *in-plane* thermal transport in vdW layered materials has received significant attention, motivated in part by potential applications in heat spreading.[4,5]

In contrast, fundamental aspects of *cross-plane* thermal transport remain relatively underexplored, despite its relevance to nanoelectronics and energy harvesting applications. For example, self-heating plays a key role in limiting the performance of field effect transistors (FETs) made of 2D materials.[6,7] While some studies have characterized heat flow at single vdW interfaces,[8–12] very little is known about the physics of "intrinsic" cross-plane thermal transport across multiple vdW layers in layered thin films. Achieving a better understanding of this is critical to realizing the potential of 2D electronics, as previous work on multi-layer MoS$_2$ transistors has shown enhancements in device mobility with increasing channel thickness (up to ~10s of nm).[13,14] In such devices, charge screening and large inter-layer electrical resistance can lead to the localization of current within the top few layers,[13] such that the dissipated heat must flow across multiple vdW interfaces before entering the substrate. It is therefore essential to understand the thickness dependence and fundamental limits of cross-plane thermal transport in vdW layered solids, particularly in materials like MoS$_2$.

A key quantity that determines thermal transport in the cross-plane direction of a material is the range of phonon mean free paths (MFPs) that carry heat. A simple estimate of the gray MFP ($\Lambda_z$) can be made using the kinetic theory, $\kappa_z \sim (1/3)\, C v_z \Lambda_z$: for MoS$_2$, using a heat capacity[15] $C \sim 2$ MJm$^{-3}$K$^{-1}$, the average sound velocity of cross-plane acoustic modes[16] $v_z \sim 2400$ ms$^{-1}$, and the cross-plane bulk conductivity[3,17] $\kappa_z \sim 2$ to $5$ Wm$^{-1}$K$^{-1}$, gives a MFP of around 1.5 to 4 nm, which corresponds to a thickness of 2 to 6 layers. A similar calculation for graphite gives a gray MFP estimate of around 3 nm, corresponding to 9 layers. This would imply that size effects (i.e. thickness dependence of $\kappa_z$) should be negligible for films thicker than ~10 nm, i.e. that the cross-plane thermal conductivity should be constant in this thickness regime. However, recent molecular dynamics (MD) simulations[18] and experimental measurements of $\kappa_z$ in graphite[19,20] have suggested surprisingly long c-axis MFPs, on the order of ~100s of nm.

These studies motivate the following key questions: (1) Are long c-axis phonon MFPs a general feature of other vdW layered systems, like the transition metal dichalcogenides (TMDs) such as $MoS_2$? (2) Can experimental observations of long cross-plane phonon MFPs in vdW materials be explained by first-principles calculations? Density functional theory (DFT) has recently proven to be very effective in understanding fundamental aspects of thermal transport in covalently bonded systems like Si,[21] but similar studies are lacking for vdW layered solids, especially quantitative comparisons with cross-plane thermal measurements. (3) What is the impact of cross-plane ballistic transport, and related size effects on the thermal resistance of thin-film TMD devices? For monolayers it is understood that interfaces dominate cross-plane thermal transport.[8,10,11] However, the transition from interface-dominated to bulk-like transport across multi-layer TMDs remains unclear to date.

In response, here we probe the spectrum of heat-carrying c-axis phonon MFPs in $MoS_2$, a vdW layered semiconductor. Through time-domain thermoreflectance (TDTR)[22] measurements of the thickness-dependent cross-plane thermal conductivity in single-crystalline films, we show that the c-axis phonon MFPs are at least ~10s of nm long, significantly larger than kinetic theory estimates. Using first-principles DFT calculations we uncover that nearly 80% of the heat at room temperature is carried by phonons with MFPs in the range 10 to 500 nm. Furthermore, we show that by suitably defining a characteristic thermal length scale, our thickness-dependent $\kappa_z$ data (with film thickness $t$ ranging from 20 nm to 240 nm) are consistent with TDTR data on bulk $MoS_2$ crystals reported previously[3,17,23] (with thermal penetration depth $d_p$ ranging from 200 nm to 1 μm). Taken together, we find good agreement between the combined data set and DFT predictions over a broad spectrum of thermal length scales, from 20 nm to 1 μm. Finally, using our measured values of the metal/$MoS_2$ and $MoS_2$/substrate interface resistances, we estimate the impact of cross-plane ballistic phonon transport on the total thermal resistance of multi-layer $MoS_2$ devices. These calculations reveal that contrary to what is typically assumed, the total thermal resistance of few nanometer thick films is not entirely interface-dominated; the lower limit is set by the ballistic resistance across the thickness of MoS$_2$, which is estimated to be ~10 m$^2$KGW$^{-1}$.

**Experimental procedure**

Single crystalline MoS$_2$ films were exfoliated onto SiO$_2$ (90 nm) on p-doped Si substrates using micromechanical exfoliation. Exfoliation yielded several MoS$_2$ films of different thicknesses on a single ~1 cm$^2$ chip. Suitable films were identified using optical microscopy, and their thicknesses measured using atomic force microscopy (AFM). An ~80 nm thick Al transducer was patterned and deposited onto the samples using electron-beam (e-beam) lithography and e-beam evaporation respectively, for TDTR measurements (see sample schematic in Figure **1**a, and Methods section). We also patterned Al onto bare regions of the SiO$_2$/Si substrate adjacent to the MoS$_2$ during the same evaporation step. This allowed us to perform reference measurements of the SiO$_2$ next to each set of samples, and helped calibrate the accuracy and consistency of our setup.

Thermal transport measurements were made using TDTR, which is a well-established optical pump-probe technique capable of measuring thermal transport in thin films and across interfaces. Details of this technique and our setup have been described previously.[22] In these experiments, the pump beam was modulated at frequencies $f_{mod}$ = 4 and 10 MHz. We used a high magnification 50× objective lens that produced a focused root mean square (rms) spot size (1/e$^2$ diameter) of $w_0$ ≈ 3 µm. An integrated dark-field microscope helped locate the samples under the laser spots.[24] Since some of the samples have lateral dimensions as small as 15 µm (especially for the thinnest films), it is important to position the laser spot well between the edges of the flake. To do this, a precision two-axis translation stage was used to map out the TDTR signal and probe beam reflectivity over the area of the sample at a fixed delay time (see Figure **1**b,c). A spot was chosen at the center of the sample within a region where the TDTR lock-in voltages and probe reflectivity were uniform, and TDTR scans were taken at that location with pump-probe delay times of 100 ps to 3.7 ns. The analysis scheme discussed below was used to simultaneously fit the normalized in-phase signal $V_{in}$, and the *ratio* (= -$V_{in}$/$V_{out}$), to a three-dimensional (3D) heat diffusion model that considers anisotropic transport.[1]

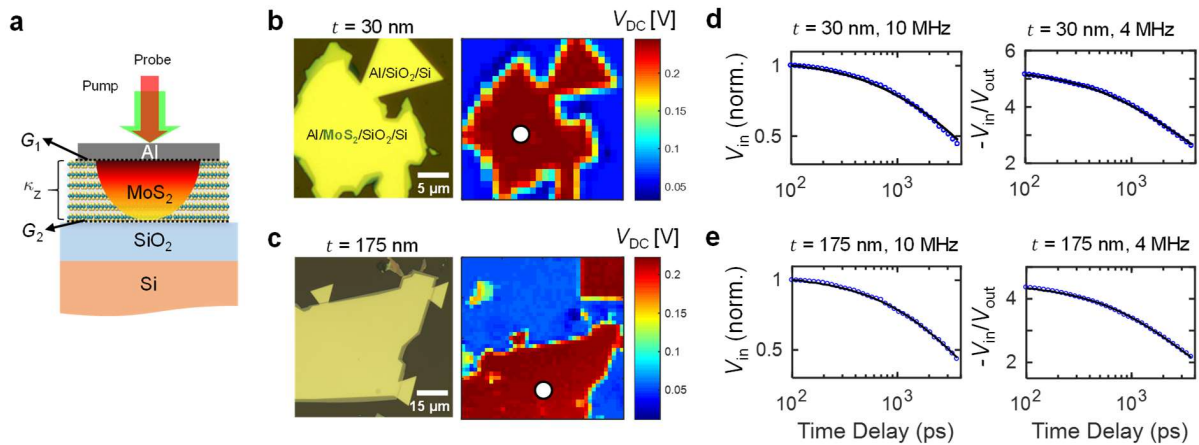

**Figure 1.** (a) Cross-sectional schematic of samples under study, showing the three unknown parameters: $G_1$ (TBC between Al and MoS$_2$), $G_2$ (TBC between MoS$_2$ and SiO$_2$), and $\kappa_z$ (cross-plane thermal conductivity of MoS$_2$). The pump (green, 532 nm) and probe (red, 1064 nm) lasers used for TDTR are shown schematically. (b) and (c) Optical micrographs and probe reflectivity maps (taken at a fixed delay time of 0 ps) for 30 nm and 175 nm thick MoS$_2$ samples, respectively. Probe reflectivity maps are used to locate a uniform region away from the sample edges, where TDTR measurements are taken (white circles). (d) and (e) TDTR data (symbols) and best fits (lines): normalized in-phase signal $V_{in}$ at $f_{mod}$ = 10 MHz and $-V_{in}/V_{out}$ ratio at $f_{mod}$ = 4 MHz for 30 nm and 175 nm thick films, respectively.

The sample stack consists of Al/MoS$_2$/SiO$_2$/Si (see Figure 1a). The thicknesses of Al and MoS$_2$ were measured using AFM, while the SiO$_2$ thickness was characterized using ellipsometry to be 90 ± 1 nm. All measurements were performed at room temperature. The thermal conductivity of Al was estimated using in-plane electrical conductivity measurements and the Wiedemann-Franz law, $\kappa_{Al} \approx 170$ Wm$^{-1}$K$^{-1}$. The thermal conductivity of the p-type Si substrate, and volumetric specific heat of Al, SiO$_2$, MoS$_2$ and Si were taken from literature.[15,25–28] To reduce the uncertainties associated with slight variations in the laser spot size between measurements on different samples, reference data were taken on the Al/SiO$_2$/Si regions next to each flake. Adjustments were made in the spot size (<5 % variation across samples) to keep the fitted SiO$_2$ thermal conductivity fixed at 1.4 Wm$^{-1}$K$^{-1}$. No $f_{mod}$ dependence was observed in the thermal conductivity of SiO$_2$ and TBC of the Al/SiO$_2$ interface ($\approx$ 130 MWm$^{-2}$K$^{-1}$) for modulation frequencies between 4 and 10 MHz.

In the MoS$_2$ sample stack, there are four unknown parameters for each sample thickness $t$. They are the intrinsic cross-plane and in-plane thermal conductivities of the MoS$_2$ layer, $\kappa_z$ and $\kappa_r$, and the TBCs at the Al/MoS$_2$ and MoS$_2$/SiO$_2$ interfaces, $G_1$ and $G_2$, respectively. The in-plane thermal conductivity is held fixed at $\kappa_r$ = 90 Wm$^{-1}$K$^{-1}$, based on prior measurements of bulk MoS$_2$ crystals by Liu et al.[3] Here, the authors had measured a spot-size dependent $\kappa_r$, likely due to the partial exclusion of ballistic phonons with in-plane mean free paths larger than the spot diameter. Our estimate for $\kappa_r$ is obtained by linearly interpolating their data to an rms spot diameter of 3 μm. To simplify our analysis we assume that $\kappa_r$ is independent of $t$, at least within the range of thicknesses (20 nm < $t$ < 240 nm) measured here. This is consistent with previous arguments by Minnich[29] and Gu et al.[30] This assumption is further discussed below.

This assumption leaves three unknown parameters for each sample: $\kappa_z$, $G_1$ and $G_2$. To extract a unique value for $\kappa_z$, we use a combination of $V_{in}$ and *ratio* (=-$V_{in}$/$V_{out}$) signals, at two different modulation frequencies, 4 MHz and 10 MHz. This tandem fitting approach is similar to that used by Meyer et al.[31] and is supported by our sensitivity analysis (see Supporting Information Section 1). For films with $t$ < 150 nm, we first estimate $G_1$ by fixing $\kappa_z$ and $G_2$, and fitting the in-phase signal $V_{in}$ (normalized at +100 ps) at the higher $f_{mod}$ of 10 MHz. Next, fixing $G_1$ at this value, the *ratio* data at the lower $f_{mod}$ of 4 MHz are fit for $\kappa_z$ and $G_2$. This process is repeated until the values of $\kappa_z$, $G_1$ and $G_2$ each change by less than 1% between successive iterations. We verify that the final fit results are not sensitive to the choice of initial values. For films with $t$ >150 nm, measurement sensitivity to bottom interface TBC, $G_2$, is relatively low. For these, we follow the same procedure as above, except that $G_2$ is held fixed at 21 ± 5 MWm$^{-2}$K$^{-1}$ based on the thin film results, further discussed below. Our methodology is generally similar to that used by Zhang et al.[20] and Jang et al.[32] for thickness-dependent $\kappa_z$ measurements of graphite and black phosphorus, respectively. Error bars are calculated by propagating uncertainties in the assumed thermophysical parameters, mainly the Al thickness (± 1 nm) and rms laser spot size (± 2%), and for the thick films also $G_2$ (± 5 MWm$^{-2}$K$^{-1}$).

We note that a recent experimental study[33] reported thickness-dependent in-plane thermal conductivity of MoS$_2$ films in the range 2.4 to 37.8 nm. To check whether this thickness-dependence might affect our extraction of $\kappa_z$, we also analyzed our data using $\kappa_r$ estimated from these results. For the 20 and 34 nm thick films, the resulting change in $\kappa_z$ is only ~2 % and ~12 %, respectively. These uncertainties are within the experimental error bars; this further confirms that our assumption of constant $\kappa_r$ for all films does not affect the extracted trend of $\kappa_z$ versus $t$.

**Results and Discussion**

Representative TDTR data and model best fits for 30 and 175 nm thick samples are shown in Figures **1**d and **1**e, respectively. Figure S2 shows the extracted top and bottom interface TBCs, $G_1$ and $G_2$ versus $t$. The MoS$_2$/SiO$_2$ TBCs fall within a narrow range of 16 to 26 MWm$^{-2}$K$^{-1}$, in reasonable agreement with Raman thermometry measurements of monolayer MoS$_2$ on SiO$_2$ (14 ± 4 MWm$^{-2}$K$^{-1}$) by Yalon et al.[8,9] The Al/MoS$_2$ TBCs are in general higher than MoS$_2$/SiO$_2$ TBCs and also show a larger spread from 30 to 80 MWm$^{-2}$K$^{-1}$ with no systematic trend as a function of $t$. This larger variability in $G_1$ could be a result of varying degrees of surface cleanliness after the e-beam patterning process that is used to define the Al transducer.

Figure **2**a plots the extracted cross-plane thermal conductivity $\kappa_z$ as a function of layer thickness $t$; $\kappa_z$ for the thickest film ($t$ = 240 nm) is 2.0 ± 0.3 Wm$^{-1}$K$^{-1}$. This decreases with decreasing film thickness down to 0.9 ± 0.2 Wm$^{-1}$K$^{-1}$ for $t$ = 20 nm, over a two-fold reduction. Such a dependence of $\kappa_z$ on $t$ is indicative of quasi-ballistic c-axis phonon transport, and suggests that the dominant heat-carrying vibrational modes have MFPs of at least ~10s of nm. We note that $\kappa_z$ appears to saturate for the three thickest films. As discussed further below, we posit that this occurs due to the finite thermal penetration depth of the TDTR measurement.

Our measured $\kappa_z$ values for the thickest films are close to two prior measurements of bulk MoS$_2$ by Liu et al.[3] and Muratore et al.[23] who obtained $\kappa_z$ of ~2 Wm$^{-1}$K$^{-1}$, and ~2.5 Wm$^{-1}$K$^{-1}$, respectively, using a TDTR modulation frequency of 9.8 MHz.[34] However, these two bulk results are significantly lower than recent measurements by Jiang et al.,[17] who obtained a bulk $\kappa_z$ ~4.8 Wm$^{-1}$K$^{-1}$. In addition, our first-principles DFT calculations (described later), obtain a bulk value of $\kappa_z$ ~5 Wm$^{-1}$K$^{-1}$, which is in good agreement with the experimental result of Jiang et al.[17] and a recent DFT calculation by Lindroth et al.[16] that predicted $\kappa_z$ ~5.1 Wm$^{-1}$K$^{-1}$.

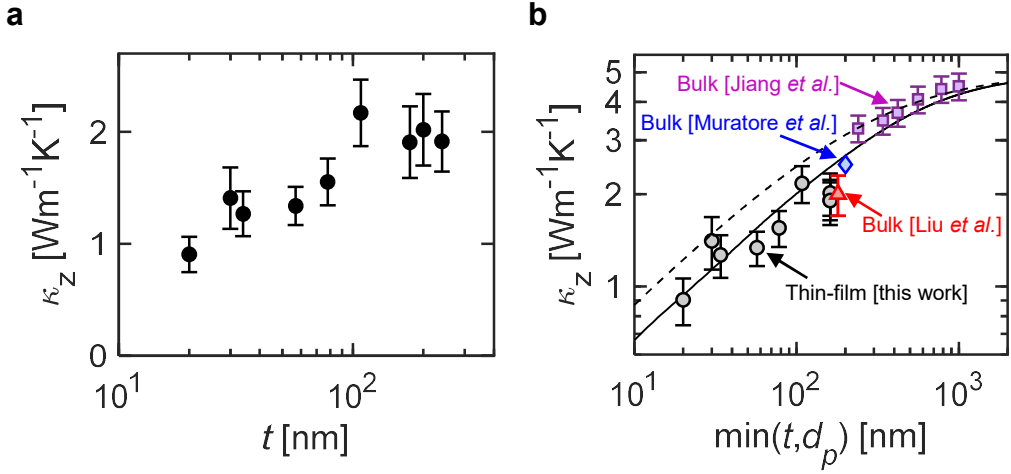

**Figure 2.** (a) Measured intrinsic cross-plane thermal conductivity $\kappa_z$ versus film thickness $t$. (b) Experimental data plotted as a function of the characteristic thermal length scale, which is the smaller of the thickness ($t$) and thermal penetration depth ($d_p$). For our measurements, $d_p \approx 160$ nm in the three thickest films at $f_{mod}$ = 4 MHz. Also shown are prior TDTR measurements of bulk MoS$_2$ by Liu et al.[3] (red triangle), Muratore et al.[23] (blue diamond), and Jiang et al.[17] (magenta squares). Solid and dashed lines are predictions of first-principles DFT calculations, with suppression functions based on the BTE[20] (Eqn. 2) and Matthiessen's rule (Eqn. 3), respectively.

To understand possible reasons behind the apparent discrepancy among the different bulk $\kappa_z$ measurements and first-principles calculations, we consider the characteristic thermal length scale (i.e. length scale over which the temperature gradient occurs) in the experiments. For TDTR measurements made at a frequency $f_{mod}$, this is determined by the thermal penetration depth[35] $d_p$, which is approximately $\sqrt{\kappa_z/\pi C f_{mod}}$. To calculate $d_p$ accurately, we solve the full 3D heat diffusion equation in the 2-layer Al/MoS$_2$ stack (see Supporting Information Section 3).

For the case of Liu et al.[3] and Muratore et al.[23], this gives $d_p$ ~180 nm (for $\kappa_z$ ~2 Wm$^{-1}$K$^{-1}$), and $d_p$ ~200 nm (for $\kappa_z$ ~2.5 Wm$^{-1}$K$^{-1}$), respectively, at $f_{mod}$ = 9.8 MHz. Jiang et al.[17] performed TDTR measurements of $f_{mod}$-dependent $\kappa_z$ in bulk MoS$_2$, and observed a reduction in the apparent $\kappa_z$ from 4.5 to 3.3 Wm$^{-1}$K$^{-1}$ while increasing $f_{mod}$ from 1 to 10 MHz. These results were interpreted based on a two-channel model that considers non-equilibrium effects between low and high-frequency phonons that have different thermal conductivities and heat capacities. The interpretation of $f_{mod}$-dependent $\kappa_z$ in modulated opto-thermal measurements has been the topic of much recent discussion.[17,25,35–40] While the treatment of near-interfacial phonon non-equilibrium deserves further attention, a first order approximation is that the contributions to heat transport of long MFP phonons with $\Lambda_z > d_p$ are suppressed at high $f_{mod}$, thereby lowering the measured apparent $\kappa_z$. This simplification is reasonable for low thermal conductivity solids with relatively broad MFP spectra as was discussed recently in the context of black phosphorus by Sun et al.,[41] and applied quantitatively to low thermal conductivity semiconductor alloys by Koh et al.[35]

In this simplified picture, we replot our thin-film $\kappa_z$ data, along with the bulk data of Liu et al.,[3] Muratore et al.[23] and Jiang et al.[17], against the thermal characteristic length scale (the smaller of $t$ and $d_p$) as shown in Figure 2b. For our thin-film samples, $d_p$ is calculated using a numerical solution of the 3D heat diffusion equation in the 4-layer stack (Al/MoS$_2$/SiO$_2$/Si). For most of our films, $d_p$ is larger than $t$. However, for the three thickest films $d_p \approx 160$ nm, which is smaller than $t$ (= 175, 200, 240 nm). Details of these calculations are provided in the Supporting Information Section 3. Following this procedure, a combined data set is obtained, where $\kappa_z$ increases from ~0.9 Wm$^{-1}$K$^{-1}$ to ~5 Wm$^{-1}$K$^{-1}$ for thermal length scales ranging from 20 nm to 1 µm. This analysis suggests that one possible reason for the discrepancy between different bulk measurements[3,17,23] could be the dependence of $\kappa_z$ on modulation frequency, and the finite thermal penetration depth. However, given that the source of MoS$_2$ crystals is typically geological, one cannot entirely rule out differences in sample quality between the various studies as contributing to the observed $\kappa_z$ variations.

Also shown in Figure 2b is a prediction of $\kappa_z$ from first-principles calculations (described below), where the effect of finite thickness is incorporated with a suppression function calculated by the Boltzmann transport equation[20] (BTE) and Matthiessen's rule. These predictions show reasonably good agreement with the combined data set over the full range of characteristic thermal length scales, from 20 nm to 1 µm. The data are thus consistent with theoretical predictions of very long c-axis phonon MFPs, and a broad spectral distribution of vibrational modes.

**First-principles DFT calculations**

To gain insight into fundamental aspects of phonon transport processes in MoS$_2$, we perform first-principles DFT calculations in the local density approximation of the exchange and correlation functional. We compute the frequency- and MFP-resolved $\kappa_z$ of MoS$_2$ by solving the phonon BTE with an iterative self-consistent algorithm.[42] Further details are provided in the Methods section and Chen et al.[42]

Calculated phonon dispersion curves for 2H-MoS$_2$ are shown in Figure 3a, which are in good agreement with experimental data.[43] Figures 3b,c plot the calculated phonon relaxation times and MFPs as a function of phonon frequency. The MFP accumulation function $\kappa_{accum}$ is calculated as a cumulative integral of the contributions to the total thermal conductivity of phonons with MFPs smaller than a certain value, and is plotted in Figure 3d. From these calculations, we infer that more than 50% of the heat at room temperature is carried by phonons with MFPs exceeding 200 nm, and nearly 80% is carried by MFPs in the range 10 nm to 500 nm. In comparison, in silicon,[21] 80% of the heat at room temperature is carried by phonons with MFPs between 40 nm and 10 μm.

Based on the MFP accumulation function, we calculate the cross-plane thermal conductivity of a film of thickness $t$ as follows,

$$\kappa_z(t) = \int_0^\infty S(Kn_\omega)\, \kappa_{partial}(\Lambda_{z,\omega})\, d\Lambda_{z,\omega} = \int_0^\infty \frac{1}{t} N(Kn_\omega)\, \kappa_{accum}(\Lambda_{z,\omega})\, d\Lambda_{z,\omega} \quad (1)$$

where $\kappa_{partial}(\Lambda_{z,\omega})$ is the MFP partial contribution function, $Kn_\omega = \Lambda_{z,\omega}/t$ is the Knudsen number, $S(Kn_\omega)$ is the heat flux suppression function, $\kappa_{accum}(\Lambda_{z,\omega}) = \int_0^{\Lambda_{z,\omega}} \kappa_{partial}(\Lambda_{z,\omega})\, d\Lambda_{z,\omega}$ is the MFP accumulation function, and $N(Kn_\omega) = -dS(Kn_\omega)/dKn_\omega$. Two cases are considered for the suppression function $S(Kn_\omega)$: one is based on a solution to the BTE for cross-plane heat flow in an anisotropic film inspired by the Fuchs-Sondheimer model[20] (Eqn. 2), and the other is based on Matthiessen's rule (Eqn. 3):

$$S_{BTE}(Kn_\omega) = 1 - Kn_\omega \left(1 - \exp\left(-\frac{1}{Kn_\omega}\right)\right) \qquad (2)$$

$$S_{Matth.}(Kn_\omega) = \frac{1}{1 + Kn_\omega}. \qquad (3)$$

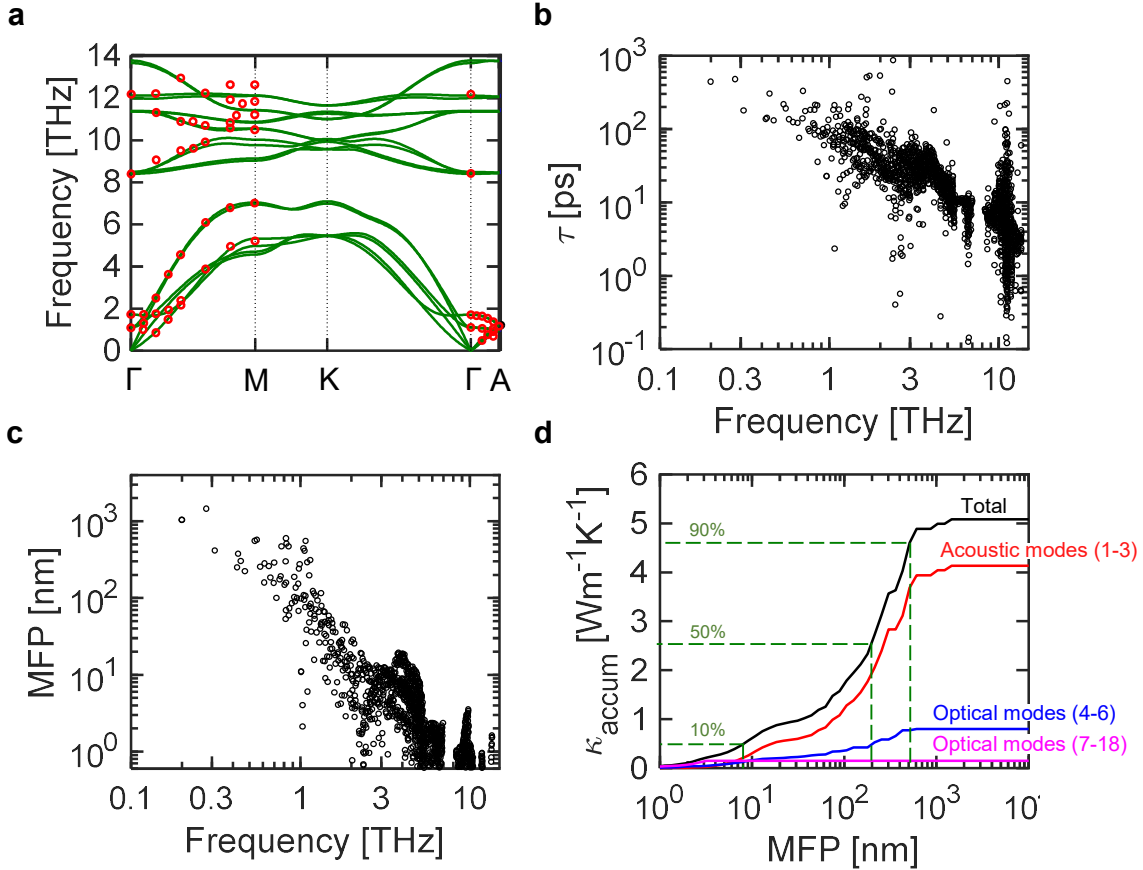

**Figure 3.** First-principles DFT calculations of $MoS_2$. (a) Phonon dispersion relations along high-symmetry directions in the Brillouin zone. Also shown with red circles are experimental data from neutron scattering on bulk $MoS_2$ crystals.[43] (b) Calculated phonon relaxation times, and (c) mean free paths (MFPs), plotted versus phonon frequency. (d) MFP accumulation function obtained from DFT calculations. The black curve is the total cross-plane thermal conductivity; the red, blue, and magenta curves are the contributions of the acoustic modes (branches 1-3, from low to high frequency on the dispersion relation), the three lowest-lying optical modes (branches 4-6), and the remaining higher-frequency optical modes (branches 7-18). Green dashed lines indicate MFPs corresponding to 10, 50 and 90 % of the accumulated total thermal conductivity.

These are plotted as solid and dashed lines, respectively, in Figure **2**b and show good agreement with the experimental $\kappa_z$ data over a large range of characteristic thermal length scales. These results have important implications for the design of thermoelectric devices based on vdW materials, as they suggest that cross-plane heat conduction can be suppressed significantly by the incorporation of defects along the c-axis,[42,44,45] such as intercalants and rotationally mismatched layers. The large phonon MFPs predicted and experimentally confirmed here offer a route to high-efficiency thermoelectrics based on nanostructuring of layered 2D materials along the c-axis.

**Implications for 2D device thermal characteristics**

To understand the impact of cross-plane ballistic phonon transport on thermal characteristics of thin-film MoS$_2$ electronic and optoelectronic devices, in Figure **4**a we plot the volumetric cross-plane thermal resistance $R_{MoS2} = t/\kappa_z$, the combined interface resistance (Al/MoS$_2$ and MoS$_2$/SiO$_2$) $R_{int.} = 1/G_1 + 1/G_2$, and the total thermal resistance $R_{total} = R_{MoS} + R_{int.}$, as a function of thickness $t$. This simplification assumes that the total resistance can be decomposed into the separate interfacial and volumetric contributions even though a large fraction of phonons undergoes quasi-ballistic transport across the thickness of the MoS$_2$ film. This assumption, which is also inherent to our data analysis methodology, is consistent with the approach commonly followed in literature when dealing with sub-continuum heat conduction across thin films.[20]

We find that $R_{MoS2}$ decreases with decreasing thickness but does not go to zero in the limit of zero $t$. This is a direct consequence of quasi-ballistic phonon transport and the diffusive scattering of long MFP phonons at the Al/MoS$_2$ and MoS$_2$/SiO$_2$ interfaces. In Figure **4**a, we also plot the calculated MoS$_2$ volumetric resistance as a function of thickness, based on DFT predictions of the phonon MFPs and the BTE suppression function (Eqn. 2). Because of ballistic transport across the film thickness, $R_{MoS2}$ saturates at a finite value of ~10 m$^2$KGW$^{-1}$ in the limit of 2 to 3 monolayers. In the absence of quasi-ballistic effects, i.e. in the diffusive regime, $R_{MoS2}$ would have been significantly lower and become vanishingly small in the monolayer limit.

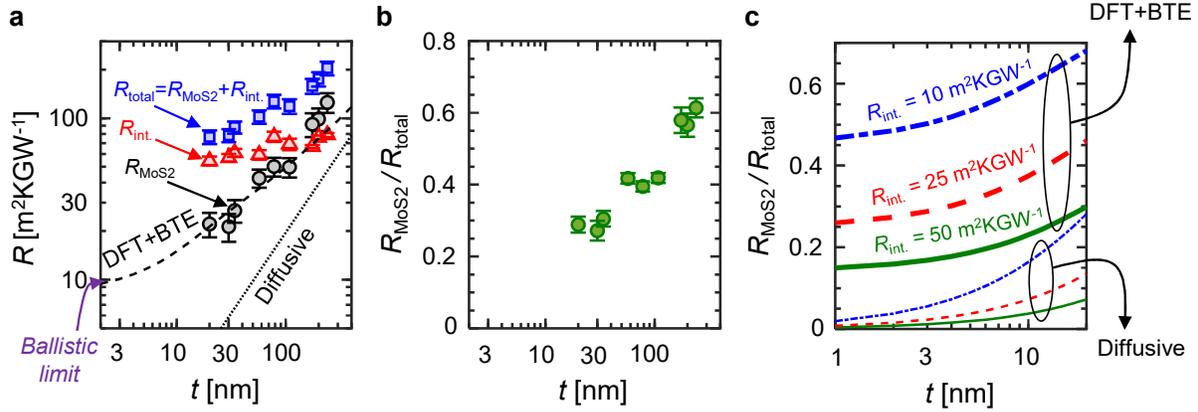

**Figure 4.** (a) Cross-plane thermal resistance of the MoS$_2$ film $R_{MoS2} = t/\kappa_z$ (black circles), combined thermal resistance of Al/MoS$_2$ and MoS$_2$/SiO$_2$ interfaces $R_{int.} = 1/G_1 + 1/G_2$ (red triangles), and total thermal resistance $R_{total} = R_{MoS} + R_{int.}$ (blue squares), plotted versus film thickness $t$. The dashed line is the calculated quasi-ballistic $R_{MoS2}$ based on first principles MFPs (Figures **3**c,d) and BTE suppression function (Eqn. 2), while the dotted line is the corresponding diffusive calculation assuming a constant $\kappa_z = 5.1$ Wm$^{-1}$K$^{-1}$. The y-intercept of the dashed line denotes the intrinsic cross-plane resistance in the ballistic limit ≈ 10 m$^2$KGW$^{-1}$. (b) Fractional thermal resistance of the MoS$_2$ film compared to the total resistance (= $R_{MoS2}/R_{total}$) plotted versus $t$. (c) Calculated fractional resistance contributed by the MoS$_2$ film for sub-20 nm thicknesses. The quasi-ballistic (heavy lines) and diffusive case (light lines) are calculated in a manner similar to (a). Three cases are shown, corresponding to $R_{int.}$ = 10, 25, and 50 m$^2$KGW$^{-1}$ in the blue dash-dotted, red dashed, and green solid lines, respectively.

An important consequence of quasi-ballistic effects is that the total thermal resistance is not dominated entirely by the interfaces, even for thin films. In Figure **4**b we plot the fractional contribution of the volumetric MoS$_2$ resistance to the total device resistance (= $R_{MoS2}/R_{total}$) versus $t$. In our experiments, for the thinnest film ($t$ = 20 nm), $R_{MoS2} \approx 22$ m$^2$KGW$^{-1}$ and $R_{total} \approx 76$ m$^2$KGW$^{-1}$, i.e. nearly 28% of the total thermal resistance is due to the volumetric component. In the diffusive limit with constant $\kappa_z = 5.1$ Wm$^{-1}$K$^{-1}$, this value is only 7% (i.e. the interfaces contribute 93%) for a 20 nm thick film. From a metrology perspective, an important consequence of this effect is our ability to experimentally measure the intrinsic component $\kappa_z$ separately from the interface resistances, down to films as thin as 20 nm. If thermal transport were to remain diffusive, the volumetric resistance component would have been too small compared to the interface resistances, and we would have not been able to extract it uniquely using TDTR.

To estimate the contribution of quasi-ballistic transport to heat flow across thin devices ($t <$ 20 nm), we plot the fractional MoS$_2$ volumetric component for different interface resistance values ($R_{int.}$ = 10, 25, 50 m$^2$KGW$^{-1}$), as shown in Figure **4**c. As before, size effects are considered by calculating the thickness-dependent $\kappa_z$ using the BTE suppression function described in Eqn. 2 above. In the extreme scaling limit of 1 to 2 nm thick films (2 to 3 monolayers), we estimate this fractional contribution to be as large as ~15 %, ~25 % and ~50 % for $R_{int.}$ = 50, 25 and 10 m$^2$KGW$^{-1}$, respectively. This suggests that even if interface quality (and TBCs) of the metal/MoS$_2$ and MoS$_2$/substrate interfaces were to be improved substantially ($R_{int.} \to 0$), cross-plane heat transport would likely still be limited by the ballistic resistance ($= \lim_{t \to 0} t/\kappa_z$) of the MoS$_2$ film, which is ~10 m$^2$KGW$^{-1}$. Note that this analysis assumes a 3D phonon dispersion for thin films, which may face its limits when the thickness becomes comparable to the phonon wavelengths (see Supporting Information Section 4). Nevertheless, this raises interesting questions about the nature of heat conduction across few-layer thick vdW layered materials, where it is often assumed that interfaces dominate the total cross-plane thermal resistance.[46] Given that the existence of long MFP *c*-axis thermal phonons has been experimentally demonstrated in graphite,[19,20] and predicted theoretically in other TMDs such as WS$_2$ and WSe$_2$,[16] the above argument may be applicable to a large class of ultra-thin vdW layered devices (see Supporting Information Section 5).

**Conclusions**

In summary, we reported thickness-dependent cross-plane thermal conductivity measurements of crystalline films of layered MoS$_2$. The cross-plane thermal conductivity shows a strong dependence on thickness in the range 20 to 240 nm, revealing quantitative information about the distribution of phonon MFPs along the *c*-axis. Combining our results with previous measurements[3,17,23] of $\kappa_z$ in bulk MoS$_2$ (made at different modulation frequencies) allowed us to map a large portion of the *c*-axis MFP spectrum from ~20 nm to ~1 µm. DFT calculations (with no fitting parameters) were able to obtain the thickness-dependent thermal conductivity over a large range of MFPs, illustrating the predictive power of first-principles phonon calculations for vdW layered materials.

Importantly, our results show for the first time that diffuse scattering of long MFP phonons imposes a lower limit on the cross-plane thermal resistance of vdW layered thin films. This can have significant implications for the thermal management of multi-layer 2D electronics[13,47] and optoelectronics (e.g. photovoltaics) where thermal transport *across* the device thickness imposes the primary bottleneck for heat dissipation. Finally, the quantitative knowledge of thermal phonon MFPs obtained here will enable the design of new applications that require engineering of the phonon spectrum. For example, the substantial contribution of long MFP phonons to $\kappa_z$ suggests that the introduction of disorder and defects along the *c*-axis can drastically suppress cross-plane thermal transport,[45] without significantly affecting electronic transport. This could have exciting implications for cross-plane thermoelectrics made of layered 2D materials,[48] potentially enabling next-generation energy harvesting and electronics cooling technologies.

**Methods**

*Sample preparation*: Flakes of $MoS_2$ were mechanically exfoliated from bulk crystals (SPI Supplies) onto 90 nm $SiO_2$ on p-type Si substrates (0.001 to 0.005 Ω.cm) using a thermal release tape (Nitto-Denko Revalpha). Samples were cleaned with an acetone/2-propanol soak and subsequently annealed in $Ar/H_2$ at 400 °C for 40 mins. This was followed by spin coating a double layer of electron-beam (e-beam) resist PMMA 495K A2/950K A4 (Microchem). The metal transducer (nominally 80 nm Al) was patterned by e-beam lithography (Raith 150, 10 kV) and deposited through e-beam evaporation. Lift-off was performed in acetone at 50 °C.

*Ab initio calculations*: First principles phonon calculations of $2H\text{-}MoS_2$ were carried out in the local density approximation (LDA) of the exchange and correlation functional using the Quantum-Espresso package.[49,50] Norm-conserving pseudopotentials were used to approximate core electrons.[51] Kohn-Sham wave functions were expanded on a plane wave basis set (cut-off = 100 Rydberg). Integration of the electronic properties over the first Brillouin zone was performed using 10×10×4 Monkhorst-Pack meshes of *k*-points[52]. Structural and cell relaxations were performed by a Broyden-Fletcher-Goldfarb-Shanno (BFGS) quasi-Newton algorithm with a strict convergence criterion of $10^{-8}$ Rydberg/Bohr for maximum residual force component.

Phonon dispersion relations were computed by density functional perturbation theory (DFPT),[53] with 10×10×4 $q$-point mesh (see Figure **3**a). The computed dispersion curves agree well with neutron diffraction data for bulk MoS$_2$.[43] For the calculation of lattice thermal conductivity, anharmonic third order interatomic force constants (IFCs) are also necessary besides the harmonic second order IFCs. Third order anharmonic force constants were computed by finite differences for a supercell,[54] which is a 5×5×1 replica of the unit cell and contains 150 atoms, with an interaction cut-off of 7 Å, including interactions up to the tenth shell of neighbors. Translational invariance of the anharmonic force constants was enforced using the Lagrangian approach.[54] With the second and third order IFCs, the thermal conductivity of MoS$_2$ was computed by solving the phonon BTE with an iterative self-consistent algorithm, using the ShengBTE code,[54] considering phonon-phonon and isotopic scattering. Convergence was checked with $q$-point grids up to 45×45×11. Further details are provided in Chen *et al*.[42]

## ASSOCIATED CONTENT

**Supporting Information**.

TDTR sensitivity analysis. Thermal boundary conductance (TBC) measurements of Al/MoS$_2$ and MoS$_2$/SiO$_2$ interfaces. Calculations of thermal penetration depth. Phonon wavelength contributions to cross-plane thermal conductivity. Literature survey of cross-plane thermal resistance of few-layer graphene and thin-film graphite.

## AUTHOR INFORMATION

**Corresponding authors**

*aditsood@alumni.stanford.edu

*epop@stanford.edu

**Present Addresses**

†Stanford Institute for Materials and Energy Sciences, SLAC National Accelerator Laboratory, Menlo Park, CA 94025, USA.

¶NG Next Basic Research Laboratory, Northrop Grumman Corporation, Redondo Beach, CA 90278, USA.

**Author contributions**

A.S., F.X. and E.P. conceived the project. A.S. designed and performed the TDTR measurements, analyzed experimental data, developed the theoretical model based on DFT calculations, and wrote the manuscript with input from E.P.; F.X. fabricated the samples; S.C. performed the DFT calculations; R.C. and F.L. provided inputs on data analysis; M.A., Y.C., D.D., K.E.G., and E.P. supervised the project. ‡These authors contributed equally.

**Notes**

The authors declare no competing financial interests.

## ACKNOWLEDGEMENT

We acknowledge the Stanford Nanofabrication Facility (SNF) and Stanford Nano Shared Facilities (SNSF) for enabling device fabrication, funded under National Science Foundation (NSF) award

ECCS-1542152. This work was supported in part by the NSF Engineering Research Center for Power Optimization of Electro Thermal Systems (POETS) with cooperative agreement EEC-1449548, by NSF EFRI 2-DARE grant 1542883, by AFOSR grant FA9550-14-1-0251, and by the Stanford SystemX Alliance. F.X. and Y.C. were partially supported by the Department of Energy, Office of Basic Energy Sciences, Division of Materials Science and Engineering, under contract DE-AC02-76SF00515.

# Supporting Information:

# Quasi-Ballistic Thermal Transport Across MoS$_2$ Thin Films

*Aditya Sood*[1,2,†,‡,*], *Feng Xiong*[3,‡], *Shunda Chen*[4], *Ramez Cheaito*[2], *Feifei Lian*[1,¶], *Mehdi Asheghi*[2], *Yi Cui*[5,6], *Davide Donadio*[4,7], *Kenneth E. Goodson*[2,5], *Eric Pop*[1,5,8,*]

[1]Department of Electrical Engineering, Stanford University, Stanford, CA 94305, USA. [2]Department of Mechanical Engineering, Stanford University, Stanford, CA 94305, USA. [3]Department of Electrical and Computer Engineering, University of Pittsburgh, Pittsburgh, PA 15261, USA. [4]Department of Chemistry, University of California, Davis, CA 95616, USA. [5]Department of Materials Science and Engineering, Stanford University, Stanford, CA 94305, USA. [6]Stanford Institute for Materials and Energy Sciences, SLAC National Accelerator Laboratory, Menlo Park, CA 94025, USA. [7]Ikerbasque, Basque Foundation for Science, E-48011 Bilbao, Spain. [8]Precourt Institute for Energy, Stanford University, Stanford, CA 94305, USA. [†]Present address: Stanford Institute for Materials and Energy Sciences, SLAC National Accelerator Laboratory, Menlo Park, CA 94025, USA. [¶]Present address: NG Next Basic Research Laboratory, Northrop Grumman Corporation, Redondo Beach, CA 90278, USA. [‡]Equal contribution.

[*]Corresponding authors: aditsood@alumni.stanford.edu, epop@stanford.edu

# 1. TDTR sensitivity analysis

To determine TDTR measurement sensitivity to the different parameters of interest, we calculate the sensitivity coefficients $S_\alpha$ as follows:

$$S_\alpha = \frac{\partial \log(Signal)}{\partial \log(\alpha)}$$

where *signal* could either refer to the normalized *in-phase voltage* ($V_{in}$) or the *ratio* (= - $V_{in}/V_{out}$), and the parameter $\alpha$ could be the cross-plane thermal conductivity $\kappa_z$, the Al/MoS$_2$ thermal boundary conductance (TBC) $G_1$, or the MoS$_2$/SiO$_2$ TBC $G_2$. These are plotted in Figure S1 for a 20 nm thick film (a, b), and a 200 nm thick film (c, d).

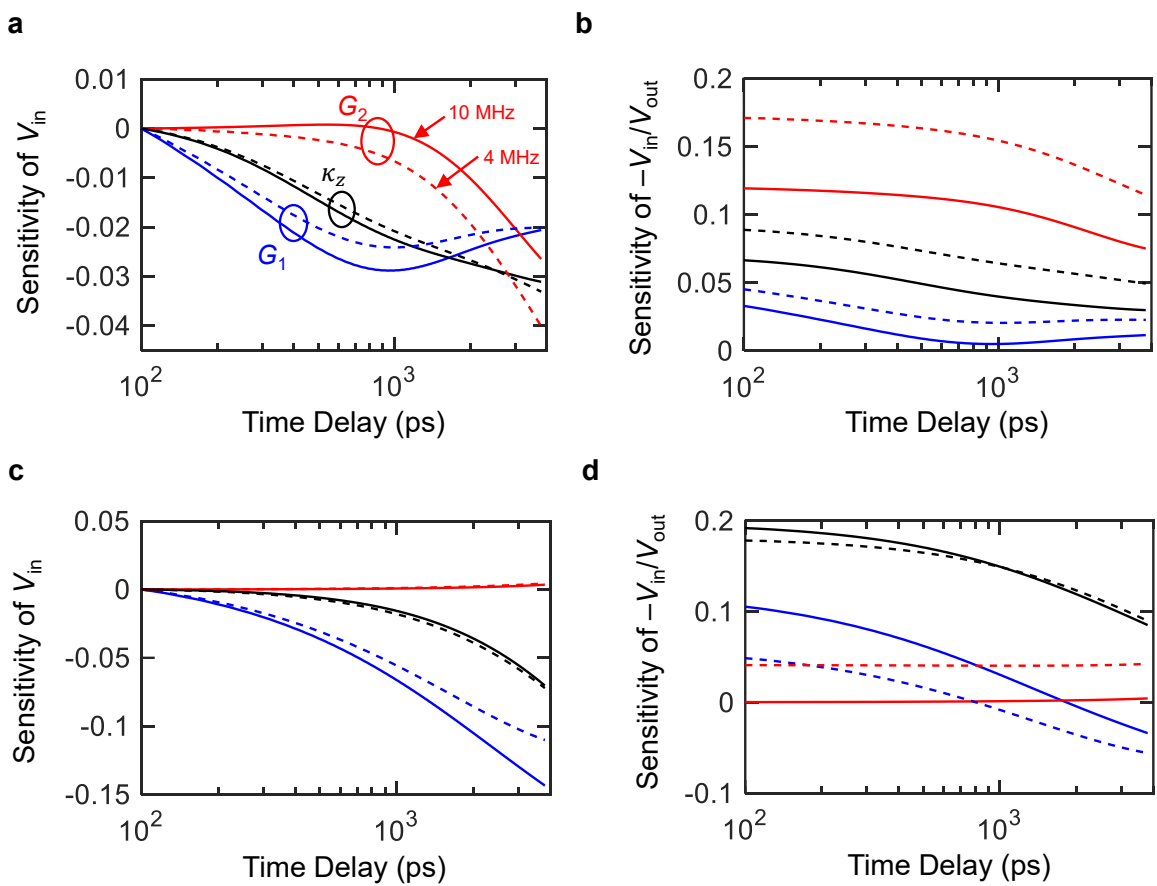

**Figure S1.** Sensitivity coefficients plotted for (a),(b): $t$ = 20 nm, $G_1$ = 70 MWm$^{-2}$K$^{-1}$, $G_2$ = 25 MWm$^{-2}$K$^{-1}$, $\kappa_z$ = 0.9 Wm$^{-1}$K$^{-1}$, and (c),(d): $t$ = 200 nm, $G_1$ = 34 MWm$^{-2}$K$^{-1}$, $G_2$ = 21 MWm$^{-2}$K$^{-1}$, $\kappa_z$ = 2 Wm$^{-1}$K$^{-1}$. Legend: black ($\kappa_z$), blue ($G_1$), red ($G_2$). Solid lines (10 MHz), dashed lines (4 MHz).

## 2. Thermal boundary conductance (TBC) measurements

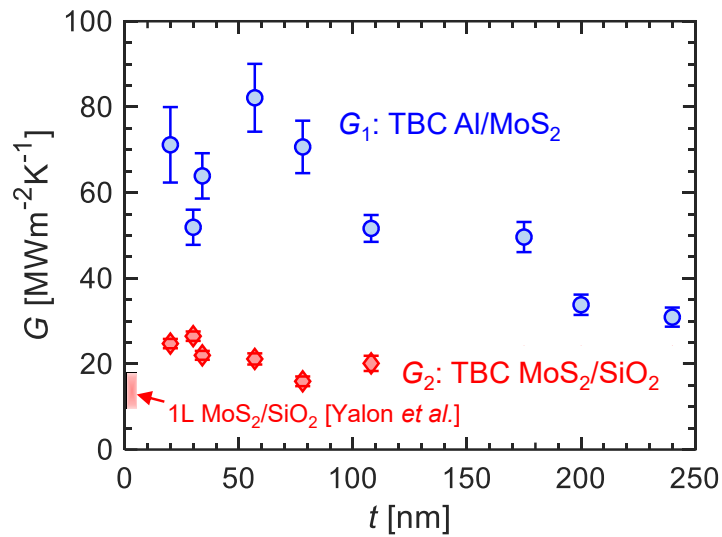

**Figure S2.** Al/MoS$_2$ ($G_1$) and MoS$_2$/SiO$_2$ ($G_2$) TBCs plotted versus film thickness $t$, shown by the blue circles and red diamonds, respectively. Also shown for comparison are TBC measurements between monolayer MoS$_2$ and SiO$_2$ obtained by Raman thermometry[1,2] (red shaded region represents the error bars of the reported result).

## 3. Thermal penetration depth calculations

To calculate the thermal penetration depth ($d_p$) in the TDTR measurements, we solve the full 3D heat diffusion equation in the multilayer stack. This is solved in the frequency domain under a sinusoidal heat flux excitation using methods described elsewhere[3,4]. We compute the amplitude of temperature oscillations $\Delta T(r,z)$ at the modulation frequency $f_{mod}$; $d_p$ is the distance from the top surface at which $\Delta T(r,z)$ is reduced to $1/e$ of its maximum value.

Figure S3(a) shows $\Delta T(r,z)$ within a 300 nm thick MoS$_2$ film – this case is representative of one of the thick samples measured in our study (for which $\kappa_z \sim 2$ Wm$^{-1}$K$^{-1}$). The simulation is carried out on a multilayer stack of Al/MoS$_2$/SiO$_2$/Si using a 4-layer model. The thermal properties of the various layers are provided in the main text. The TBCs of the Al/MoS$_2$ and MoS$_2$/SiO$_2$ interfaces are 40 MWm$^{-2}$K$^{-1}$ and 20 MWm$^{-2}$K$^{-1}$ respectively, although these do not affect $d_p$ significantly. The heat flux is modulated at $f_{mod} = 4$ MHz, since this is the frequency at which we extract $\kappa_z$. Note that $d_p$ is affected both by $f_{mod}$ and the laser spot diameter ($w_0$); in these simulations, $w_0 = 3$ μm. Figure S3(b) plots $\Delta T(z)$ at $r = 0$. From this we estimate $d_p \approx 160$ nm.

The same procedure is used to calculate $d_p$ for the bulk samples measured in previous studies[5–7] using a 2-layer model (Al/MoS$_2$). In each case, the simulations are performed using the reported $\kappa_z$, $f_{mod}$ and $w_0$ values. A representative calculation[5] is shown in Figures S3(c),(d).

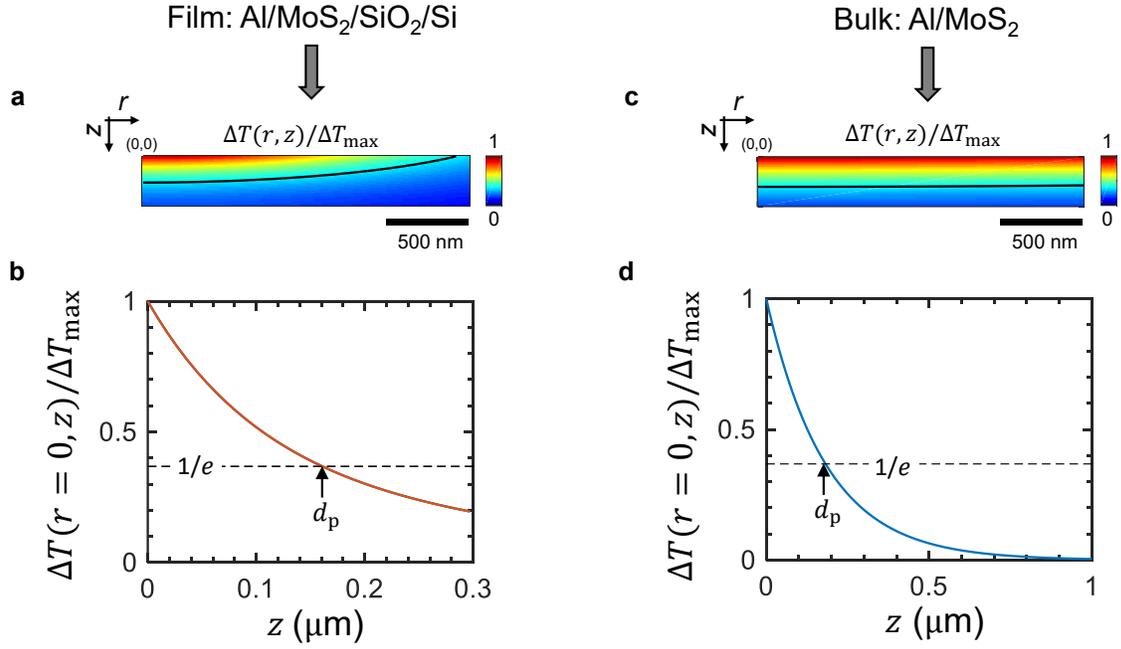

**Figure S3.** (a) Normalized amplitude of temperature oscillations in a 300 nm thick MoS$_2$ film with $\kappa_z = 2$ Wm$^{-1}$K$^{-1}$, $f_{mod} = 4$ MHz, $w_0 = 3$ μm. The film is part of a multilayer stack: Al/MoS$_2$/SiO$_2$/Si, representative of the samples measured in this study. (b) Line-out along $r = 0$, with the dashed line indicating a $1/e$ thermal penetration depth of $d_p \approx 160$ nm. (c) Normalized amplitude of temperature oscillations in a bulk MoS$_2$ substrate[5] with $\kappa_z = 2$ Wm$^{-1}$K$^{-1}$, $f_{mod} = 9.8$ MHz, $w_0 = 24$ μm. (d) Line-out along $r = 0$, indicating $d_p \approx 180$ nm.

## 4. Phonon wavelength contributions to thermal conductivity

We use DFT calculations to determine the range of phonon wavelengths that contribute to thermal transport along the $c$-axis. Figure S4 shows the thermal conductivity accumulation function plotted versus wavelength at 300 K. Based on this, the median wavelength is $\lambda \sim 1.5$ nm. If we posit that the MoS$_2$ film must have a thickness of at least $\sim 3\lambda$ in order to have a '3D' phonon dispersion, we estimate a minimum thickness of $\sim 5$ nm. For $t < 5$ nm, more detailed calculations may be needed to understand the effect of confinement on phonon band structure and cross-plane thermal transport.

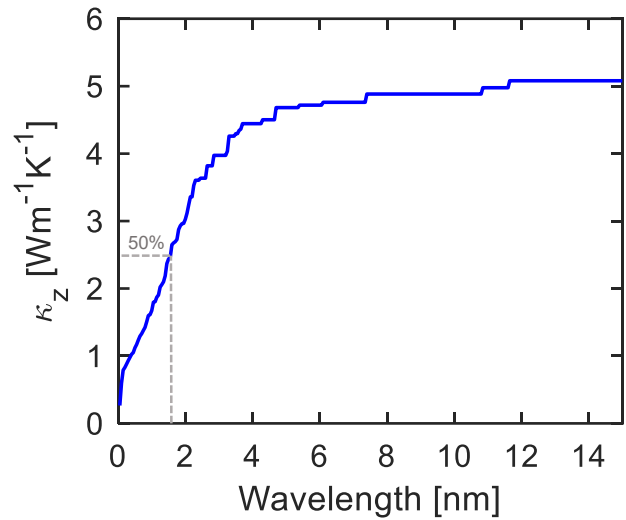

**Figure S4.** Calculated cumulative distribution function of the cross-plane thermal conductivity ($\kappa_z$) versus phonon wavelength at 300 K.

## 5. Cross-plane thermal transport in thin-film graphite and few-layer graphene

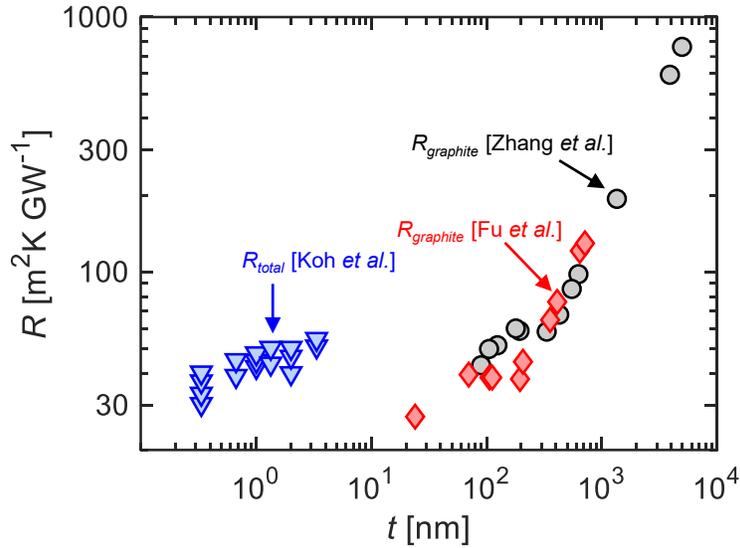

**Figure S5.** A summary of cross-plane thermal resistance measurements of crystalline graphite thin-films and few-layer graphene from literature. Intrinsic cross-plane thermal resistance measurements are from Zhang et al.[8] (90 nm < $t$ < 5 μm), shown in black circles, and Fu et al.[9] (24 nm < $t$ < 714 nm), shown in red diamonds. The intrinsic resistance is defined as $R_{\text{graphite}} = t/\kappa_z$. For the case of Fu et al.[9] this is calculated by subtracting out the estimated interface contribution. Total cross-plane thermal resistance measurements of Au/Ti/few-layer-graphene/SiO$_2$ interfaces for 0.3 < $t$ < 3 nm are from Koh et al.[10], shown as blue triangles; the total resistance including the interfacial contribution is $R_{\text{total}} = R_{\text{n−graphene}} + R_{\text{interfaces}}$. The plateau in intrinsic thermal resistance in Zhang et al.[8] and Fu et al.[9] could be related to the onset of quasi-ballistic thermal transport. A comparison to the total thermal resistance values for few-layer-graphene by Koh et al.[10] suggests that a contributing factor to the thickness-independent $R_{\text{total}}$ could be the strongly-ballistic transport of thermal phonons propagating along the *c*-axis of the thin-films.